\documentclass[a4paper,usenatbib]{mnras2}

\usepackage{times}

\usepackage{graphicx}	
\usepackage{amsmath}	
\usepackage{amssymb}	
\usepackage{amsfonts}
\usepackage{color}

\usepackage{url}
\usepackage[flushleft]{threeparttable}

\usepackage{etoolbox}

\hypersetup{colorlinks=true,linkcolor=blue,citecolor=blue,filecolor=blue,urlcolor=blue}
\makeatletter
\patchcmd{\NAT@citex}
  {\@citea\NAT@hyper@{%
     \NAT@nmfmt{\NAT@nm}%
     \hyper@natlinkbreak{\NAT@aysep\NAT@spacechar}{\@citeb\@extra@b@citeb}%
     \NAT@date}}
  {\@citea\NAT@nmfmt{\NAT@nm}%
   \NAT@aysep\NAT@spacechar\NAT@hyper@{\NAT@date}}{}{}

\patchcmd{\NAT@citex}
  {\@citea\NAT@hyper@{%
     \NAT@nmfmt{\NAT@nm}%
     \hyper@natlinkbreak{\NAT@spacechar\NAT@@open\if*#1*\else#1\NAT@spacechar\fi}%
       {\@citeb\@extra@b@citeb}%
     \NAT@date}}
  {\@citea\NAT@nmfmt{\NAT@nm}%
   \NAT@spacechar\NAT@@open\if*#1*\else#1\NAT@spacechar\fi\NAT@hyper@{\NAT@date}}
  {}{}
  
\usepackage{xspace}
  
\def\aj{AJ}

\def\apj{ApJ}
\def\apjl{ApJ}
\def\apjs{ApJS}
\def\ao{Appl.~Opt.}
\def\apss{Ap\&SS}
\def\aap{A\&A}

\def\mnras{MNRAS}

\def\pasp{PASP}

\def\rmxaa{RMxAA}

\def\arcsec{\hbox{$^{\prime\prime}$}}

\newcommand{\ionic}[2]{#1$\,${\scshape{#2}}\xspace}
\newcommand{\ionf}[2]{#1$\,${\scshape{#2}}}

\newcommand{\bc}[1]{\textrm{\footnotesize{\color{blue}#1}}}

\newcommand{\orcidicon}{\includegraphics[width=0.26cm]{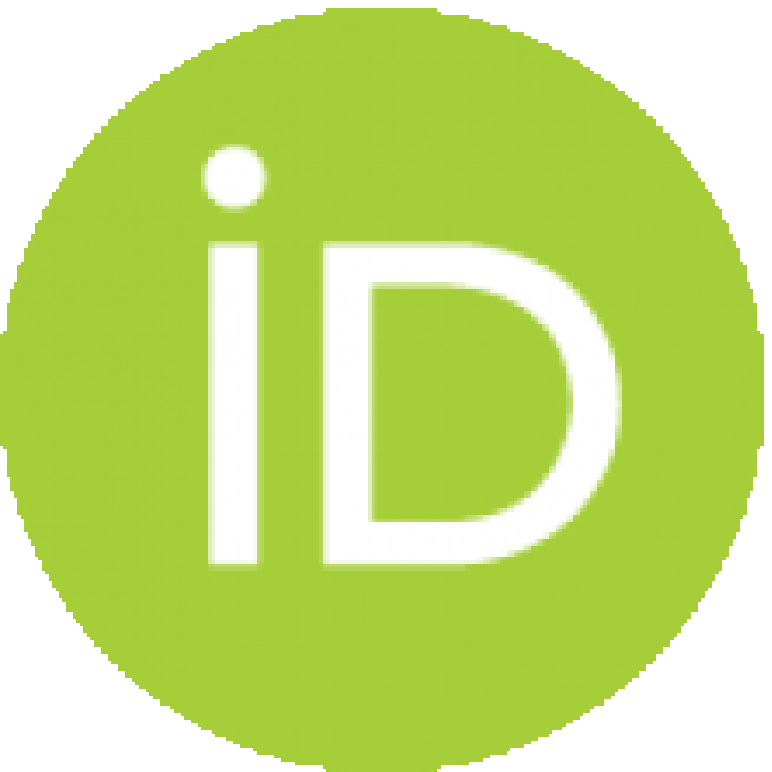}}

\newcommand{\orcidauthor}[1]{\,$^{\href{https://orcid.org/#1}{\orcidicon}}$}

\title[3D kinematic modeling of Abell 48]{3D spatio-kinematic modeling of Abell 48, a planetary nebula around a Wolf-Rayet [WN] star\thanks{Based on observations with the  Australian National University 2.3-m Advanced Technology Telescope under
program 1100147.}}

\author[A.~Danehkar]{A.~Danehkar\orcidauthor{0000-0003-4552-5997}$^{{\bc{1}}}$\thanks{E-mail: \href{mailto:danehkar@umich.edu}{danehkar@umich.edu}} 
\\
$^{1}$Department of Astronomy, University of Michigan, 1085 S. University Ave., Ann Arbor, MI 48109, USA
}

\pubyear{2021}

\begin{document}
\label{firstpage}

\pagerange{\pageref{firstpage}--\pageref{lastpage}} \pubyear{2020}

\maketitle

\begin{abstract}
The planetary nebula (PN) Abell\,48 (PN\,G029.0+00.4) is around a rare Wolf-Rayet [WN5] star whose stellar history is as yet unknown. Using the integral field observations of the H$\alpha$ $\lambda$6563 and [N\,{\sc ii}] $\lambda$6584 line emissions, we conducted a comprehensive spatio-kinematic analysis of this PN. A three-dimensional spatio-kinematic ionization model was developed with the kinematic modeling tool \textsc{shape} to replicate the observed spatially-resolved velocity channels and position--velocity diagrams. According to our kinematic analysis of the H$\alpha$ emission, this object possesses a deformed elliptic toroidal shell with an outer radius of 23\,arcsec and a thickness of 15\,arcsec associated with an integrated H$\alpha$ emission-line expansion of $\sim 35 \pm 5$ km\,s$^{-1}$, a maximum poloidal expansion of around $70 \pm 20$ km\,s$^{-1}$ at an inclination angle of $\sim 30^{\circ}$ with respect to the line of sight, and a position angle of $\sim 130^{\circ}$ measured from east toward north in the equatorial coordinate system. Furthermore, [N\,{\sc ii}] kinematic modeling reveals the presence of narrow ($\sim 3$\,arcsec) exterior low-ionization structures surrounding the main elliptical shell, which could have formed as a result of shock collisions with the interstellar medium. The torus-shaped morphology of this PN could be related to its unusual hydrogen-deficient [WN] nucleus that needs to be inspected further.
\end{abstract}




\section{Introduction}
\label{a48:introduction}

Abell\,48 ($=$ PN\,G029.0$+$00.4 $=$ A66\,48 $=$ PK\,029$+$00\,1) was detected and classified as a planetary nebula (PN) by \citet{Abell1955}. Its ring-shaped morphology was first seen in the photographic images collected for the Palomar Observatory Sky Survey \citep{Abell1966}. \citet{Kromov1968} described its morphology as a ring-like shape with a mean angular diameter of 40$\arcsec$ using images prepared for the Catalogue of Galactic Planetary Nebulae \citep{Perek1967}. In the H$\alpha$ survey of PNe with the Palomar 1.5 m telescope, \citet{Jewitt1986} identified its morphological class as a ring-type without any halo. The boundaries of this PN correspond to angular dimensions of 46$\arcsec \times$ 38$\arcsec$, i.e., a maximum radius of 23$\arcsec$ and a shell thickness of around 15$\arcsec$ when seen at roughly  $10$ percent of its mean surface brightness in the H$\alpha$ band of the SuperCOSMOS H$\alpha$ Sky Survey \citep[SHS;][]{Parker2005}. 

The central star of Abell\,48 has caused some controversies \citep{Wachter2010,Todt2013,Frew2014}. Previously, \citet{Zuckerman1986} described its stellar characteristics as [WC] with \ionic{O}{vi}. We use the square brackets to separate the Wolf-Rayet (WR) central stars of planetary nebulae from their massive WR counterparts. However, \citet{Wachter2010} classified it as a massive WN6 with a ring nebula. \citet{Depew2011} also suggested that it may be either [WN] or [WN/WC] surrounded by a PN similar to the PN PB 8 with a [WN/WC] star \citep{Todt2010}, and IC\,4663 with a [WN] star \citep{Miszalski2012}. Detailed spectral analyses of the central star of Abell\,48 pointed to either [WN5] \citep{Todt2013} or [WN5-6] \citep{Frew2014} instead of a massive WN6 (Pop I) star. More recently, a comparison between the \ionic{H}{ii} $\lambda$4686 lines emitted by the Abell\,48 central star and the WN7 star of WR\,148 exhibits an asymmetric profile toward the red wing in WR\,148, whereas we see a rather symmetric profile in Abell 48 that could be a signature of a lower mass-loss rate associated with a low-mass [WR] star \citep{Munoz2017}. 

The oxygen abundance of the PN Abell\,48 was reported to be ${\rm O}/{\rm H} = 1.6 \times 10^{-4}$ \citep{Danehkar2014} and $2.7 \times 10^{-4}$ \citep{Todt2013}. Its low metallicity may help to form its [WR] central star \citep{Zijlstra2006,Kniazev2008,MillerBertolami2011}. A self-consistent photoionization model of this object constructed by \citet{Danehkar2014} also reproduced its nebular spectrum with a low metallicity (${\rm O}/{\rm H} = 1.4 \times 10^{-4}$) at a distance of 1.9\,kpc using the hydrogen-deficient expanding model constrained by \citet{Todt2013} as its ionizing source. However, a solar metallicity (${\rm O}/{\rm H} = 4.9 \times 10^{-4}$) was also derived by adopting different physical conditions \citep{Frew2014}, which may not lead to a [WN] star. We should note that electron densities of $\sim 750$ \citep{Danehkar2014,Frew2014} and $1000$\,cm$^{-3}$ \citep{Todt2013} were also determined from the [\ionf{S}{ii}] line ratio. Moreover, an electron temperature of around $7000$\,K was also deduced from the [\ionf{N}{ii}] line ratio \citep{Todt2013,Danehkar2014,Frew2014}. Although the [\ionf{O}{iii}] $\lambda$4363 auroral line is extremely faint, higher uncertain temperatures of $\lesssim 9950$\,K \citep{Frew2014} and $11900$\,K \citep{Todt2013} were also calculated using the [\ionf{O}{iii}] $\lambda\lambda$4959,5007/$\lambda$4363 ratio. The accuracy of chemical abundances largely depends on the electron temperature assumed for empirical abundance analysis \citep[see e.g.][]{Danehkar2021a}, so this PN could have a metallicity ranging from sub-solar to solar composition.

In this work, we employ integral field spatially resolved kinematic maps and position--velocity (P--V) arrays of the PN Abell\,48 to build a three-dimensional (3D) kinematic model of this object. In Section \ref{a48:observations}, we briefly describe the observation. The kinematic observations are present in Section \ref{a48:results}, followed by a spatio-kinematic model in Section \ref{a48:kinematic}. Our conclusion is drawn in Section \ref{a48:conclusion}.

\section{Observation}
\label{a48:observations}

\begin{figure}
\begin{center}
\includegraphics[width=2.5in]{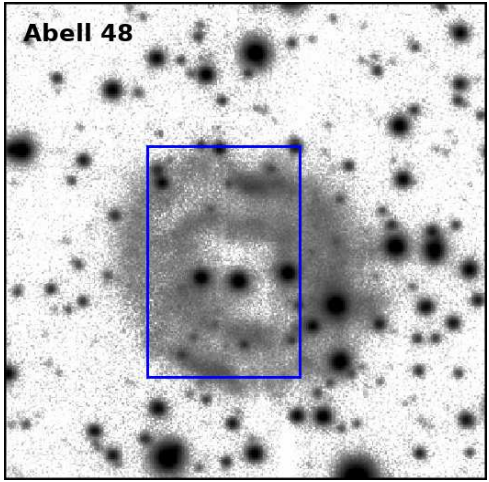}
\caption{The $r_{P1}$-band image of Abell\,48 from the PS1 survey \citep{Chambers2016,Flewelling2020}. 
The rectangle corresponds to the IFU FOV ($25\arcsec \times 38\arcsec$) of the WiFeS taken in 2010 April.
The image is oriented with north up and east to the left.
\label{a48:ps1:fig1}%
}%
\end{center}
\end{figure}

The moderate-resolution kinematic observation of the PN Abell\,48 studied in this work was collected on April 22, 2010 with 
the Wide Field Spectrograph \citep[WiFeS;][]{Dopita2007,Dopita2010}, on the 2.3-m telescope operated by the Australian National University (program number 1100147, PI: Q.A. Parker). The WiFeS is a double-beam integral field unit (IFU) spectrograph with a field-of-view (FOV) of $25\arcsec \times 38\arcsec$, a spatial resolution of $1\farcs0\times0\farcs5$, and a full width at half-maximum of the reconstructed point-spread function (PSF) of approximately $2\arcsec$ on $4096 \times 4096$-pixel CCD detectors. Volume phase holographic gratings utilized by the WiFeS yield spectral resolutions of $R \sim 3000$ and $R \sim 7000$. The observation was taken with an exposure time of 1200 sec, while the WiFeS grating was configured to $R\sim 7000$ that leads a linear wavelength dispersion per pixel of $0.45$ {\AA} in the red arm over $\lambda\lambda$5222--7070 {\AA}, resulting in a velocity channel resolution of around $21$ km\,s${}^{-1}$. 

The \textsc{iraf} pipeline \textsf{wifes} was used to perform the standard data reduction \citep[fully described in][]{Danehkar2014,Danehkar2021}. This procedure includes averaged bias subtraction, pixel-to-pixel sensitivity correction using flat-field frames, spectral calibration with Cu--Ar arc lamp exposures, spatial calibration using wire frames, atmospheric refraction correction, cosmic ray removal, and flux calibration.

Figure \ref{a48:ps1:fig1} shows the $r_{P1}$-band image ($\lambda_{\rm mean} = 6215$ {\AA} and $\Delta \lambda \approx 1600$ {\AA}) of Abell\,48 in the Pan-STARRS1 (PS1) survey \citep{Chambers2016,Flewelling2020} that was retrieved from the Mikulski Archive for Space Telescopes (MAST). The rectangle on the PS1 image depicts the WiFeS areal footprint used for the 2010 observation.

\section{Kinematic Results}
\label{a48:results}

\begin{figure}
\begin{center}
\includegraphics[width=3.5in]{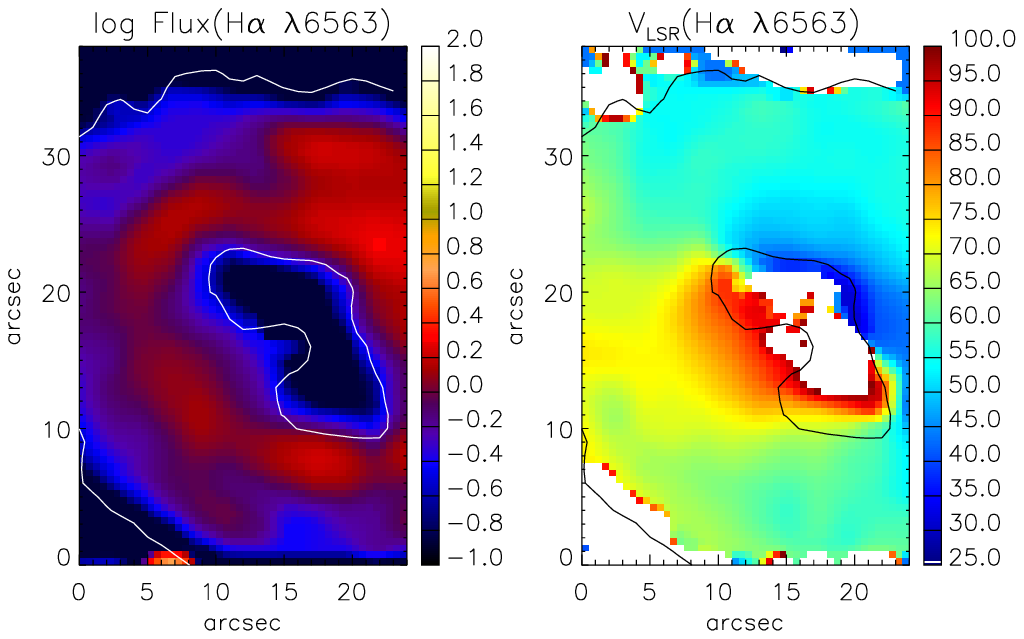}\\
\includegraphics[width=3.5in]{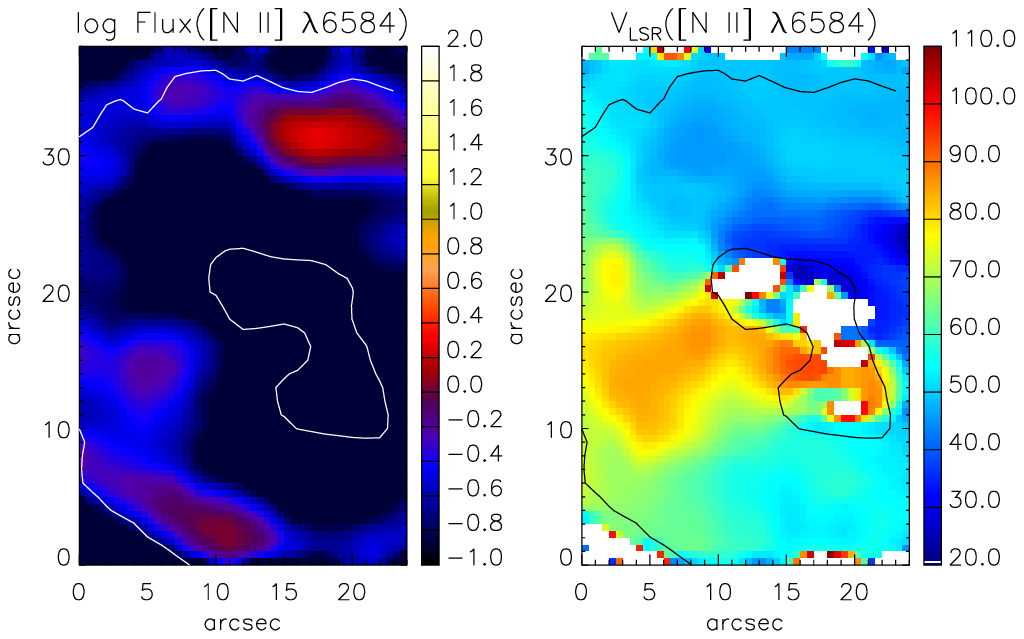}
\caption{IFU maps of Abell 48 in the H$\alpha$ $\lambda$6563\,{\AA} (top panels) and [N\,{\sc ii}] $\lambda$6584\,{\AA} emission (bottom). From left to right, the logarithmic flux map in  $10^{-15}$~erg\,s${}^{-1}$\,cm${}^{-2}$\,spaxel${}^{-1}$ unit, and LSR radial velocity map in km\,s${}^{-1}$. 
The contours correspond to $\sim 10$ percent of the mean H$\alpha$ surface brightness of this object seen in the SHS. 
The field is oriented with north up and east to the left.
\label{a48:ifu:fig2}%
}%
\end{center}
\end{figure}

Figure \ref{a48:ifu:fig2} presents the spatially resolved maps of the flux intensity and radial velocity of Abell\,48 for the H$\alpha$ $\lambda$6563 and [N\,{\sc ii}] $\lambda$6584 emission seen in the WiFeS field. The velocity map is in the local standard of rest (LSR) and was transferred using the \textsc{idl} implementation (written by D.~Nidever, 2006) of the \textsc{iraf} task \textsf{rvcorrect}. The white/black contours in the figures correspond to about $10$ percent of the SuperCOSMOS H$\alpha$ Sky Survey \citep[SHS;][]{Parker2005} mean surface brightness of this PN. It can be seen that the IFU maps illustrate an inclined ring-shaped nebula.

\begin{figure*}
\begin{center}
\includegraphics[width=6.in]{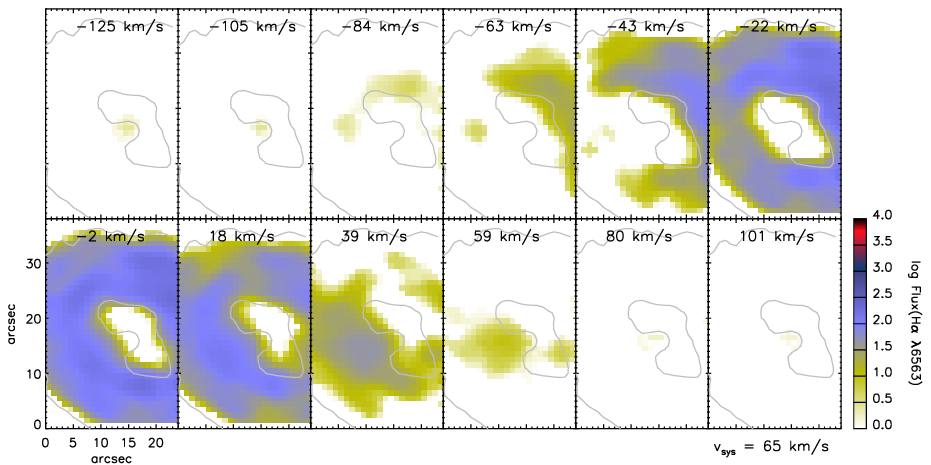}\\
\includegraphics[width=6.in]{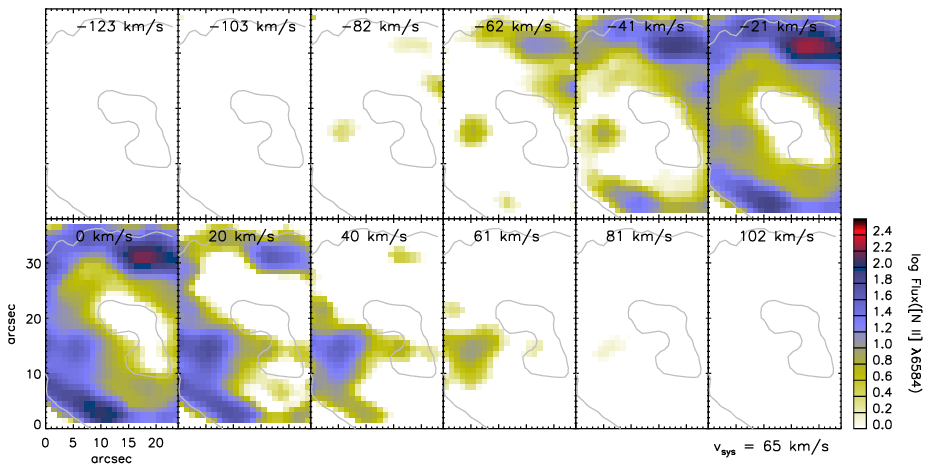}
\caption{Velocity slices of Abell\,48 along the H$\alpha$ $\lambda$6563\,{\AA} (top panels) and [N\,{\sc ii}] $\lambda$6584\,{\AA} (bottom) emission with central velocities in km\,s${}^{-1}$ at the top. The velocity width of each slice is around $21$ km\,s${}^{-1}$. 
The systemic velocity is $v_{\rm sys}=65$\,km\,s${}^{-1}$ in the LSR frame. 
The color bars correspond to the logarithmic fluxes in $10^{-15}$~erg\,s${}^{-1}$\,cm${}^{-2}$\,spaxel${}^{-1}$. 
The gray contours are associated with $\sim 10$ percent of the mean surface brightness of this PN seen in the SHS. 
The field is oriented with north up and east to the left.
\label{a48:slic:fig3}
}
\end{center}
\end{figure*}

\begin{figure*}
\begin{center}
{\footnotesize (a) Observed P--V diagrams}\\ 
\includegraphics[width=3.5in]{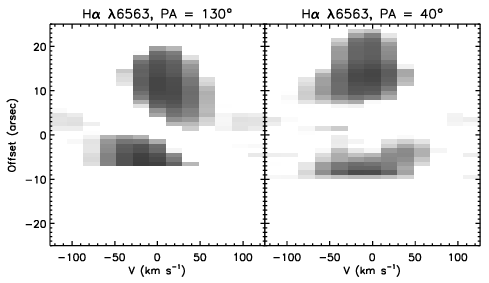}%
\includegraphics[width=3.5in]{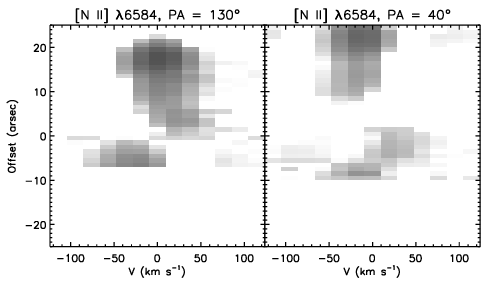}\\
{\footnotesize (b) Modeled P--V diagrams}\\ 
\includegraphics[width=3.5in]{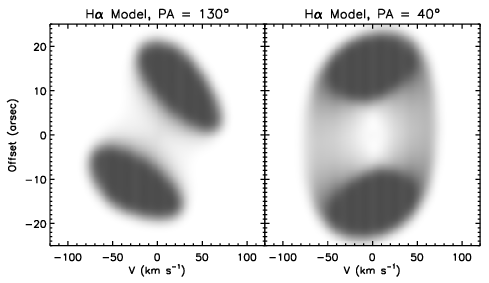}%
\includegraphics[width=3.5in]{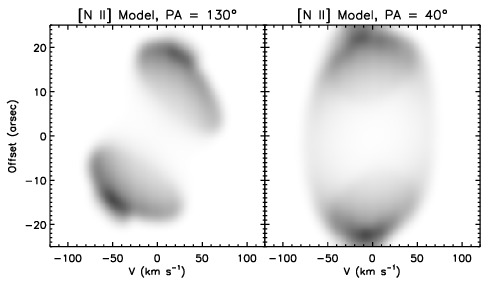}
\caption{\textit{Top Panels:} Observed P--V diagrams of Abell\,48 in the H$\alpha$ $\lambda$6563 (right panels) and [N\,{\sc ii}] $\lambda$6584  emission (left).
Slits are oriented with ${\rm PA} = 130^{\circ}$ and $40^{\circ}$ through the central star.
The velocity axis is given in km\,s${}^{-1}$ relative to the systemic velocity of the PN. The angular offset at 0 arcsec corresponds to the central star.
\textit{Bottom Panels:} The associated synthetic P--V arrays rendered from the spatio-kinematic model with density laws describing the H$\alpha$ $\lambda$6563 (right panels) and [N\,{\sc ii}] $\lambda$6584 (left) observations of Abell\,48.
\label{a48:pv:diagram}%
}
\end{center}
\end{figure*}

We obtained an expansion velocity of $35\pm5$ km\,s$^{-1}$ that corresponds to the half width at half maximum (HWHM) measurement of the H$\alpha$ $\lambda$6563 emission line in the integrated spectrum extracted from the entire nebula covered by the WiFeS FOV. The systemic velocity measured from the H$\alpha$ $\lambda$6563 emission line was found to be $v_{\rm sys} = 65\pm 5$\,km\,s$^{-1}$ in the LSR, which agrees with the previous results \citep{Todt2013,Danehkar2014}. The P--V array and velocity channel maps in the H$\alpha$ emission also point to $v_{\rm sys} = 62$\,km\,s$^{-1}$, albeit with a velocity resolution of about $21$\,km\,s$^{-1}$. The systemic velocity estimated from the emission-line profile integrated over all the spaxels of the entire nebula seems to be more accurate than the P--V diagram and velocity channels with a slice interval of $\sim 21$\,km\,s$^{-1}$.

Figure~\ref{a48:slic:fig3} presents the logarithmic flux distributions of the H$\alpha$ and [N\,{\sc ii}] emissions in a series of 12 velocity channel maps with a width of about $21$\,km\,s$^{-1}$ that span from $-125$ to $102$\,km\,s$^{-1}$. The stellar continua were determined and subtracted from them. The central velocity is also provided at the top of each channel, which is relative to the systemic velocity $v_{\rm sys}=65$ km\,s$^{-1}$. The H$\alpha$ velocity channel maps show different components in the $-43$, $-22$, $-2$, $18$, $39$ km\,s$^{-1}$ velocities, which are associated with the front and rear ends of a toroidal shell expanding at $\sim 35 \pm 5$ km\,s$^{-1}$ with respect to the central star. Similarly, the [N\,{\sc ii}] channel maps suggest the same expanding velocity, but with a high distribution of N$^{+}$ towards the outer boundary of the nebula. However, the outer radius of the nebula in the [N\,{\sc ii}] emission is about 3\,arcsec larger than that in the H$\alpha$ emission, which could suggest the presence of a narrow, low-ionization N$^{+}$ layer surrounding the photoionized H$^{+}$ main shell. Assuming that this nebula is a circle-shaped ring inclined at an angle ($i=30^{\circ}$, see \S\,\ref{a48:kinematic}) projected on the sky plane, the symmetric axis of this object has a position angle (PA) of $130^{\circ} \pm 2^{\circ}$ from the direction of the north towards the east in the equatorial coordinate system (ECS). 

Figure~\ref{a48:pv:diagram} (top panels) presents the P--V diagrams in the H$\alpha$ $\lambda$6563 (right) and [N\,{\sc ii}] $\lambda$6584  emission (left) produced from the IFU datacube for two slits positioned through the central star and oriented with ${\rm PA} = 130^{\circ}$ of the symmetric axis and $40^{\circ}$ perpendicular to the symmetric axis. The velocity axes are relative to the systemic velocity ($v_{\rm sys}=65$ km\,s$^{-1}$). The central star is located at 0 arcsec in the angular axes. The stellar continua were also determined and removed from the P--V diagrams. The two separate knots seen in the H$\alpha$ P--V diagrams are the front and rear parts of a projected torus. It can be seen that the top and bottom sides of the toroidal shell reach a deprojected poloidal velocity of $v_{\textrm{pol-exp}}=70\pm20$ km\,s$^{-1}$ with respect to the central center.  As can be seen, the IFU footprint did not fully capture the lower portion of the P--V arrays. Again, similar to the velocity channels, the nebular size seen in the [N\,{\sc ii}] P--V array at ${\rm PA}= 40^{\circ}$ is slightly larger than what we see in the H$\alpha$ diagram, implying the existence of an exterior low-ionization structure beyond the photoionized  H$^{+}$ region. This low-ionization N$^{+}$ zone could be produced by the interaction with the interstellar medium (ISM).

\section{Spatio-kinematic Modeling}
\label{a48:kinematic}

To reproduce spatially resolved kinematic maps and P--V arrays, we have employed the kinematic modeling tool \textsc{shape} v5.0 \citep[][]{Steffen2006,Steffen2011}. This tool allows us to interactively create and modify 3D geometries using polygon meshes, and to define velocity and density laws for 3D geometries. It applies a ray-casting algorithm to volumic grids of defined 3D geometries to produce radiative transfer solutions through them,  resulting in synthetic images, P--V diagrams, and velocity channels. To find the best-fitting model, it is necessary to iteratively adjust 3D geometries, velocity and density laws, orientation and inclination  angles until we get solutions that closely resemble observations. 3D geometries can also be exported from the \textsc{shape} platform to 3D models in the standard triangle language (STL) file format that can also be transformed into other appropriate file formats such as the Alias/WaveFront object file format using a mesh processing program such as MeshLab \citep{Cignoni2008} for online publication.
This modeling tool has been employed to construct 3D kinematic models of several PNe, such as NGC\,2392 \citep{Garcia-Diaz2012}, Hb\,5 and K\,3-17 \citep{Lopez2012}, Th\,2-A \citep{Danehkar2015}, Abell\,14 \citep{Akras2016}, and Hb\,4 \citep{Derlopa2019}.

\begin{figure*}
\begin{center}
{\footnotesize \hspace{2.5cm} Wireframe Model \hspace{6.2cm} Interactive 3D Model\hspace{1cm}}\\
\includegraphics[width=4.5in]{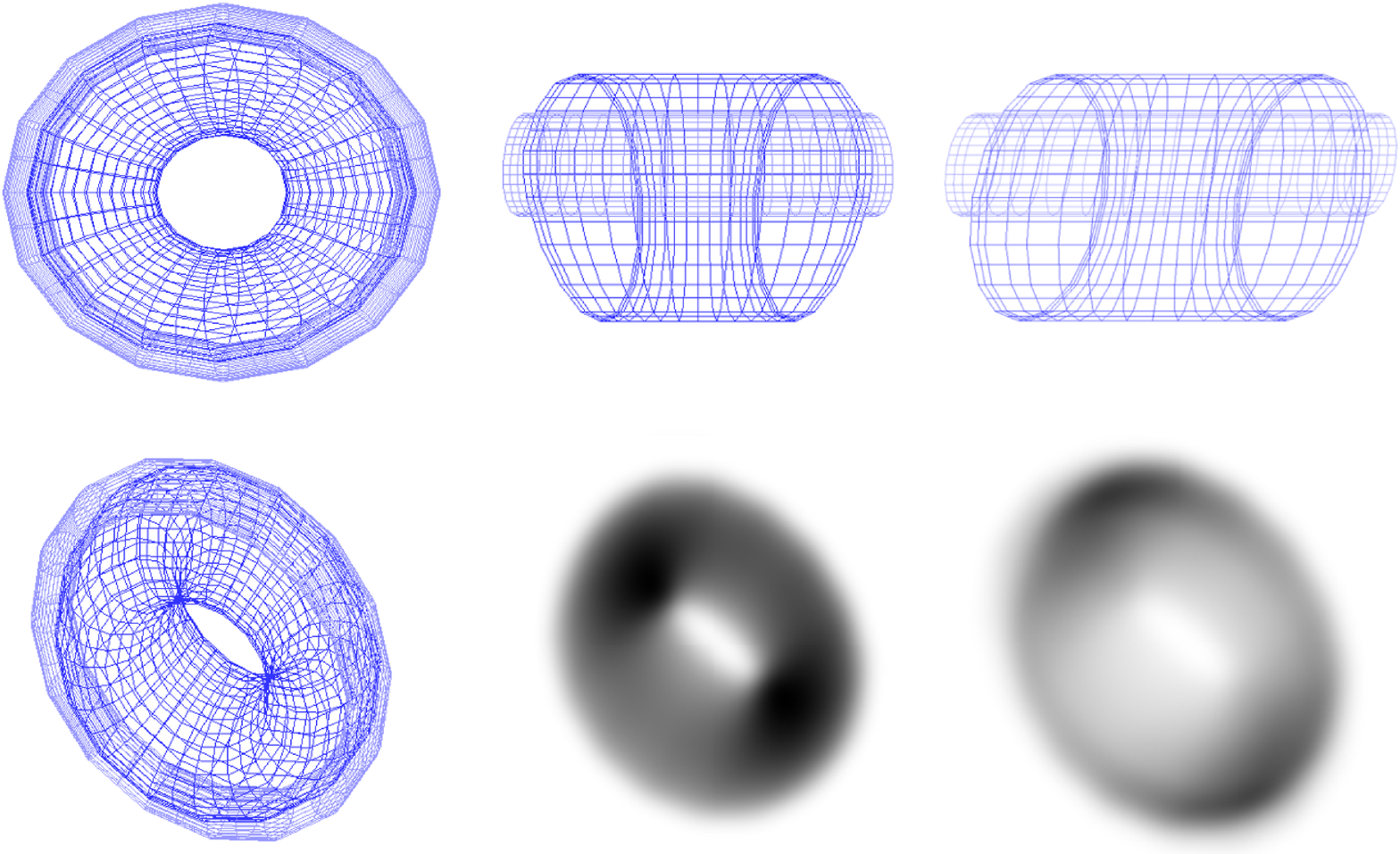}%
\includegraphics[width=2.4in]{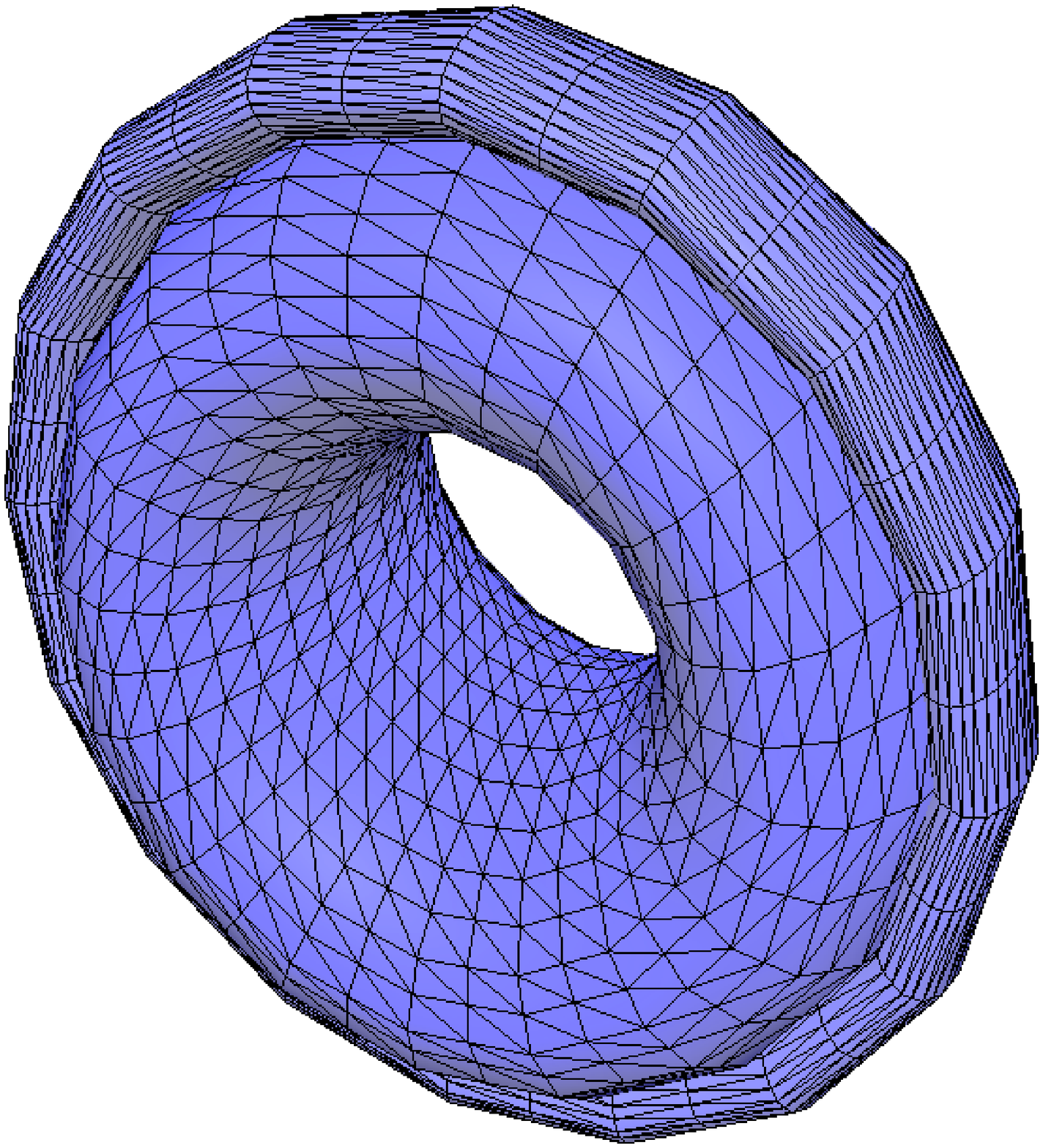}
\caption{\textit{Left Panel}: Spatio-kinematic wireframe model of Abell\,48 seen from the top view ($i=0^{\circ}$), the front and side view ($i=90^{\circ}$), and the best-fitting inclination ($i=30^{\circ}$) and orientation (${\rm PA} = 130^{\circ}$), followed by the rendered images for the H$\alpha$ $\lambda$6563 and [N\,{\sc ii}] $\lambda$6584 emissions made with two different density laws, respectively.
The exterior elliptic torus is used only for modeling the [N\,{\sc ii}] emission.
\textit{Right Panel}: Interactive 3D model available on Sketchfab (\href{https://skfb.ly/o7nxA}{https://skfb.ly/o7nxA}) for the online journal.
\label{a48:shape:fig5}
}
\end{center}
\end{figure*}

Figure \ref{a48:shape:fig5} (left) depicts the kinematic wireframe model of Abell 48 at three different views (top: $i=0^{\circ}$, front and side: $90^{\circ}$), as well as their best-fitting inclination ($i=30^{\circ}$) and orientation (${\rm PA} = 130^{\circ}$) that well reproduce the IFU observations, followed by the two rendered images associated with two different density laws for the H$\alpha$ and [N\,{\sc ii}] emissions, respectively. The 3D model is also provided in the Alias/WaveFront object file format on Sketchfab with a representative image in Figure \ref{a48:shape:fig5} (right), and archived on Zenodo.\footnote{\href{https://doi.org/10.5281/zenodo.5511247}{https://doi.org/10.5281/zenodo.5511247}}The mesh model reproducing the H$\alpha$ emission is built of a torus that is deformed using the size, shear, and squish modifiers in the \textsc{shape} platform. The \textit{size} modifier transforms it into an elliptic torus by stretching it by a factor of 2 along the $z$ axis. Additionally, this elliptic torus is also surrounded by an exterior elliptic torus that reproduces the observed P--V array in the [N\,{\sc ii}] emission, but the outer torus is deactivated for the H$\alpha$ kinematic model via the density function. This kinematic model helps to reproduce a maximum poloidal velocity of $v_{\textrm{pol-exp}}=70$ km\,s$^{-1}$ with respect to the center, while the elliptic toroidal shell expands with an averaged integrated H$\alpha$ emission-line velocity of $v_{\textrm{HWHM}}=35$ km\,s${}^{-1}$. Moreover, the \textit{shear} operator pushes the geometry by misaligned pressure in the $x$ direction, which reproduces the pattern seen in the observed P--V array along the minor axis (${\rm PA} = 40^{\circ}$). Similarly, \citet{Jones2010} applied the shear modifier to an elliptical geometry to emulate the asymmetric morphology of Abell~41. We also utilize the \textit{squish} operator to make a slightly asymmetric elliptic torus whose front side is slightly wider than its rear, resulting in a better match with the observed P--V arrays. 

The deformation in the nebular shell could be created by the ISM interaction. Moreover, the exterior narrow N$^{+}$ layer around the main H$^{+}$ shell in Abell 48 could also be owing to shock-ionization with the ISM. Similarly, a deformed ring is visible in the PN Hb\,4 \citep{Danehkar2021} that could be produced as the PN moves through the ISM. Moreover, the PN M\,2-42 was found to have a pair of asymmetric bipolar outflows \citep{Danehkar2016} that could be a sign of the ISM interaction \citep[e.g.][]{Wareing2007}. The possibility of the ISM interaction was also considered for the PN Abell~41 \citep{Jones2010}, for which asymmetric brightness and shape are found. It is worthwhile mentioning that a later outburst from the post-AGB star may also contribute to asymmetric and deformed structures as it pushes away the previously expelled shell, and also results in the formation of low-ionization shocked regions \citep[see e.g. NGC 5189 in][]{Danehkar2018}.

\begin{table}
\caption{Parameters of the spatio-kinematic model of Abell\,48.\label{a48:parameters}}
\begin{center}
\begin{tabular}{ll}
\noalign{\smallskip}
  \hline\noalign{\smallskip}
Parameter & Value \\
\noalign{\smallskip}
   \hline\noalign{\smallskip}
Inclination of the symmetric axis, $i$ & $30^{\circ} \pm 2^{\circ}$  \\
Position angle of the symmetric axis, PA 				& $130^{\circ} \pm 2^{\circ}$  \\
Outer radius of the shell, $r_{\rm out}$(H$\alpha$)			& $23\pm2$ arcsec \\
Outer radius of the shell, $r_{\rm out}$([N\,{\sc ii}])			& $26\pm2$ arcsec \\
Thickness of the shell, $\delta r$(H$\alpha$)				& $15\pm2$ arcsec \\
HWHM expansion, $v_{\textrm{HWHM}}$(H$\alpha$)		 	& $35\pm5$ km\,s${}^{-1}$\\
Maximum poloidal expansion, $v_{\textrm{pol-exp}}$	  				& $70\pm20$ km\,s${}^{-1}$\\
LSR systemic velocity, $v_{\rm sys}$(H$\alpha$)				& $65\pm5$ km\,s${}^{-1}$  \\
\noalign{\smallskip}\hline
\end{tabular}
\end{center}
\end{table}

To reproduce the P--V line profiles, a linear radial velocity law is defined as a function of the radial distance $v$[km\,s$^{-1}$]$ = 4 \times r$[arcsec]. This type of expansion is called the homologous (Hubble-type) velocity law \citep[][]{Steffen2009}. To emulate the IFU observations of the [N\,{\sc ii}] $\lambda$6584 emission, we also employ a density law in the main shell that radially increases by a power of 3 outward from the nebular center that closely resembles the N$^{+}$ ion fraction predicted by the photoionization model \citep[see Figure 7 in ][]{Danehkar2014}. Moreover, the outer shell around the main shell is also reproduced the N$^{+}$ zone in the P--V array at ${\rm PA} = 40^{\circ}$ that could be due to shock collisions with the ISM. While we model the H$\alpha$ observations, we deactivate this outer region through the density operator in \textsc{shape}. To improve the [N\,{\sc ii}] P--V  diagrams, we also modify the density profile of the main shell using a sinusoidal expression defined by the polar angle in the spherical coordinate system so that the density decreases as one moves from the equator ($\phi=90^{\circ}$) to the poles ($\phi=0^{\circ}$ and $180^{\circ}$), with a greater decrease at the south pole ($\phi=180^{\circ}$). This inhomogeneous density model is only used for the [N\,{\sc ii}] observations. Similarly, \citet{Akras2012a} defined an inhomogeneous density law with thinner polar distributions to recreate the P--V diagrams of BD\,+30$^{\circ}$3639.  

\begin{figure*}
\begin{center}
{\footnotesize (a) H$\alpha$ Model}\\ 
\includegraphics[width=6.in]{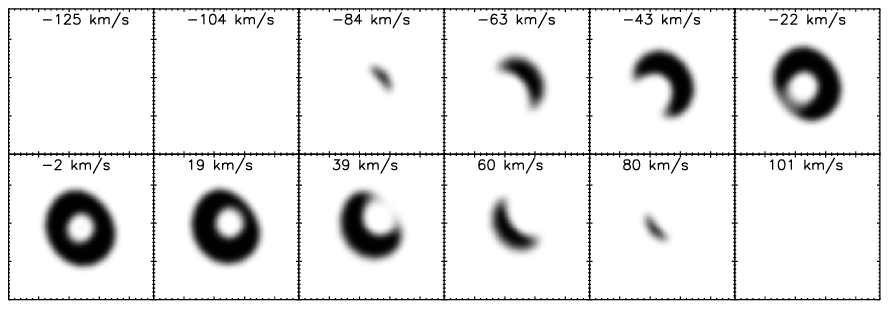}\\
{\footnotesize (b) [N\,{\sc ii}] Model}\\ 
\includegraphics[width=6.in]{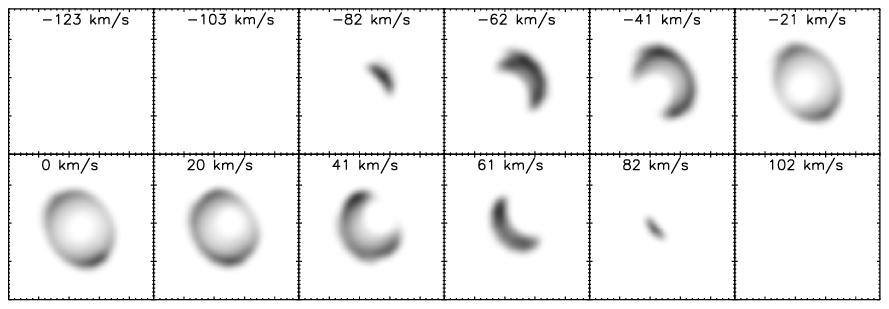}
\caption{The synthetic velocity channels rendered from the spatio-kinematic model with density laws describing the H$\alpha$ $\lambda$6563 (top panels) and [N\,{\sc ii}] $\lambda$6584 (bottom) observations of Abell\,48.
\label{a48:channel:fig5}
}
\end{center}
\end{figure*}

Figure~\ref{a48:pv:diagram} shows a comparison of the synthetic P--V arrays generated by the spatio-kinematic model (bottom panels) with the P--V diagrams observed in H$\alpha$ $\lambda$6563 and [N\,{\sc ii}] $\lambda$6584 (top panels) for two slits oriented along the symmetric axis (${\rm PA}=130^{\circ}$) and vertical to the symmetric axis (${\rm PA}=40^{\circ}$). As seen in the figure, there are reasonable agreements between the observed and modeled P--V diagrams, while the lower parts of the P--V diagrams were not completely observed by the IFU FOV. Figure~\ref{a48:channel:fig5} also presents the synthetic velocity channels of the model, which can also be compared with the spatially resolved velocity maps shown in Figure~\ref{a48:slic:fig3}. The synthetic results provide a good match to the observed maps, so the model well reproduces the kinematic properties and ionization distributions of the H$^{+}$ and N$^{+}$ ions in this object. Some discrepancies between the [N\,{\sc ii}] observations and synthetic P--V arrays are still present, which could be related to asymmetric low-ionization structures (see Figure~\ref{a48:ifu:fig2}) possibly due to the ISM interaction.

The parameters of the best-fitting models are listed in Table~\ref{a48:parameters}. The symmetric axis of this PN has ${\rm PA}=130^{\circ} \pm 2^{\circ}$ measured from east toward north in the ECS. This PA corresponds to a Galactic position angle of $192.8 \fdg 2$ from the north Galactic pole toward the Galactic east. Previously, without any constraints on velocity channel maps and P--V diagrams, a PA of $135^{\circ}$ was derived \citep{Danehkar2014}. We also see that the velocity slices and P--V arrays are well matched if the inclination of the major axis is $i=30^{\circ} \pm 2^{\circ}$, whereas $i=35^{\circ} \pm 2^{\circ}$ was found by \citet{Danehkar2014} without detailed kinematic analysis using the older version of \textsc{shape}.\footnote{Note that $i=-35^{\circ}$ reported by \citet{Danehkar2014} using \textsc{shape} v4.5 should be $i=35^{\circ}$ with the latest version.}
Assuming $D=1.6\pm 0.5$\,kpc estimated by \citet{Frew2014}, the H$\alpha$ outer radius of $23 \pm 2$ arcsec corresponds to $r_{\textrm{out}} = 0.178^{+0.074}_{-0.073}$\,pc, so
the maximum poloidal expansion $v_{\textrm{pol-exp}}=70\pm20$ km\,s${}^{-1}$ implies a dynamical age of $\tau_{\textrm{dyn}}=r_{\textrm{out}}/v_{\textrm{pol-exp}} = 2490^{+1980}_{-1220}$ yr.
As proposed by \citet{Dopita1996}, the true age could be $\tau_{\textrm{true}} = 1.5 \times \tau_{\textrm{dyn}} = 3740^{+1830}_{-2980}$ yr, in agreement with the dynamical age of 3900\,yr derived by \citet{Frew2014}.
However, \citet{Todt2013} estimated a nebular age of 6500\,yr based on a typical expansion velocity of 30 km\,s${}^{-1}$.  
Alternatively, \citet{Gesicki2000} suggested that the dynamical age should be calculated based on the expansion velocity $v_{\textrm{exp}}$ and the asymptotic giant branch (AGB) velocity $v_{\textrm{AGB}}= 0.5 \times v_{\textrm{exp}}$, so the true age of Abell\,48 could be $\tau_{\textrm{true}} = 2 r_{\textrm{out}}/(1.5 v_{\textrm{pol-exp}}) = 3320^{+1620}_{-2650}$ yr. 
The estimated dynamical ages are consistent with an ionized nebula ejected from an intermediate-mass progenitor star with a final stellar mass of $\lesssim 0.6$\,M$_{\odot}$. On the other hand, if this nebula is associated with a massive WN star,  
it should be located at 12 kpc with an ionized mass of $\sim 50$\,M$_{\odot}$ \citep{Frew2014}. 
While the expansion velocity could be related to a nebula around a massive WN star, the physical conditions and ionic abundances derived for 
this nebula are not typical of massive WN nebulae \citep{Todt2013,Danehkar2014,Frew2014}.

\section{Conclusion}
\label{a48:conclusion}

Using integral field spectroscopic observations, a comprehensive spatio-kinematic analysis of the PN Abell 48 with a [WN]-type star has been conducted that can reproduce the velocity-resolved channels and P--V arrays observed in the H$\alpha$ $\lambda$6563 and [N\,{\sc ii}] $\lambda$6584 emission line profiles. A 3D spatio-kinematic model has been constructed using a sheared elliptic torus that could be evidence of the PN-ISM interaction that happens as the nebula is expanding and moving through the ISM. The extreme poloidal expansion of $v_{\textrm{pol-exp}}=70 \pm 20$ km\,s${}^{-1}$ was replicated by velocity components of the stretched, sheared torus in the $z$ direction, while it has a mean integrated H$\alpha$ emission-line expansion of $v_{\textrm{HWHM}}=35 \pm 5$ km\,s${}^{-1}$. The line-of-sight inclination and ECS orientation were found to be $i=30^{\circ} \pm 2^{\circ}$ and ${\rm PA} = 130^{\circ} \pm 2^{\circ}$, respectively. The [N\,{\sc ii}] P--V arrays were recreated using an inhomogeneous density profile in the elliptic torus owing to the N$^{+}$ ion radial profile produced by the photoionization process \citep[see][]{Danehkar2014}, as well as a thin ($\sim 3$\,arcsec) external shell likely associated with shock ionization. Some asymmetries seen in the [N\,{\sc ii}] flux map (Figure~\ref{a48:ifu:fig2}) could also be related to shock excitation caused by the interaction with the ambient medium.

The central star of Abell\,48 has been identified as a [WN]-type star \citep{Todt2013,Frew2014}. Its hydrogen-deficient stellar characteristics may have emerged when its hydrogen-rich stellar layer was lost to a companion, but we have not yet detected any binary system in this object. We know that some PNe with similar elliptical ring-shaped morphology evolved from common envelopes with binarity, such as Abell\,63 \citep{Mitchell2007}, Abell\,41 \citep{Jones2010}, Sp\,1 \citep{Jones2012}, and HaTr\,4 \citep{Tyndall2012}, whose symmetric axes are roughly perpendicular to their binary orbital planes. An in-depth monitoring of its central star will help us assess the possibility of binarity, which could also explain the formation of its rare [WN] star and its nebular morphology.

\section*{Acknowledgements}

A.D. wishes to express gratitude to Professor Quentin Parker and the staff at Siding Spring Observatory for supporting the Australian National University 2.3-m observation and the anonymous referee for constructive comments and thoughts.



\section*{Data Availability}

The data underlying this article will be shared on reasonable request to the corresponding author. The 3D model is available on Sketchfab (\href{https://skfb.ly/o7nxA}{https://skfb.ly/o7nxA}), and archived on Zenodo (doi:\href{https://doi.org/10.5281/zenodo.5511247}{10.5281/zenodo.5511247}).







\bsp	
\label{lastpage}
\end{document}